\documentstyle[11pt]{article}
\begin{document}

\baselineskip=15pt

\thispagestyle{empty}

\vspace{20mm}

\begin{center}
{\Large 
Electroweak corrections to the $H\to t\bar{t}$ decay rate}\\[6.4ex]
{\large 
Chen Xiao-Xi, Ma Wen-Gan, Liu Yao-Yang and Sun La-Zhen}\\[2mm]
{\small Modern Physics Deptment, University of Science and
Technology of China, Anhui 230027, China.} \\[4ex]
{\large 
Chang Chao-Hsi}\\[2mm]
{\small CCAST (World Laboratory), P.O.~Box 8730, Beijing 100080, China \\
and\\
Institute of Theoretical Physics, Academia Sinica,
P.O.~Box 2735, Beijing 100080, China.\footnote{Mailing address.} }\\[20mm]
\begin{minipage}{100mm}
\centerline{\bf Abstract}
\vspace{5mm}

The electroweak corrections at one-loop level to the
process $H\rightarrow t{\bar t}$ are calculated,
especially the fresh top mass value announced recently
by CDF is concerned.
For $M_H < 1. TeV$ where a perturbative calculation is
valid, the corrections themselves would gain
a few to 20 percent increment in the decay width as the
Higgs mass $M_H$ is increasing within the
region, but they are in opposite direction to the QCD ones.
If the electroweak and QCD
corrections are concerned in the meantime,
the resultant decay width of the
mode yields a reduction about a few percent of the tree level one.
\end{minipage}
\end{center}

\vspace{20mm}
PACS number(s): 12.38.Bx; 14.80.Dq; 11.10.Gh; 12.15.Ji

\newpage

\noindent
{\Large \bf 1. Introduction}

\vskip 5mm
\begin{large}
\baselineskip 0.35in

To search for Higgs boson \cite{s1} and top-quark \cite{s2} is one of
the goals for the next generation colliders such as
CERN LEP200 and Large Hadron Collider (LHC),
as the two particles are indispensible in the nowaday promised
standard model (SM).
Even before having the colliders constructed, a tremendous
amount of energy has already been, and continues to be, devoted
to theoretical and experimental studies
on searching for signatures of the two particles.
For the Higgs boson, the four experiments at LEP-I, the
CERN $e^{+}e^{-}$ collider have
recently placed a lower direct
bound of $M_H\sim 57 GeV$ at $95\%$ C.L.\cite{s3}.
With regard to the top-quark, a direct SM lower
bound of its mass, $m_t\sim 113 GeV$, has been obtained from the
Collider Detector at Fermilab (CDF) experiments \cite{s4}.
Indirect upper bounds for $m_t$ and $M_H$ can be predicted
by the SM theory based on quantum loop phenomenology.
It is well known that heavy top-quark loop corrections to
certain low-energy and electroweak (EW) observables
(for example, the $\rho$ parameter) are proportional to
$m_t^2$, and thus the quantum effects are quite sensitive
to $m_t$. The SM consistency of all the low-energy
experimental data requires $m_t<182 GeV$ at $95\%$ C.L.,
with a center value of $m_t=125\pm 30 GeV$ \cite{s5}. However,
when we had just compiled the revised version
of this paper, a news came out that
CDF collaboration published their
new evidence and analysis on the top searching.
They suggested in their preprint that the top mass
should be as follows\cite{s21}
$$ m_t = 174\pm 10^{+13}_{-12} GeV.$$
Therefore keeping the original aspect
of the paper as much as possible, we have added
the numerical calculation on the relevant
and interest observables with the fresh top mass
in the present version (see Fig.7).

Of the loop corrections due to a heavy Higgs,
the mass dependence behaves as $\ln (M_H/M_W)^2$ in the SM.
This is the famous one-loop ``screening rule'' first
recognized by Veltman \cite{s6}.
Since the dependence of quantum loop effects on the
heavy Higgs boson is only logarithmic, low-energy observables
are not very sensitive to $M_H$.
However, it was pointed out some time ago by
Dicus and Mathur \cite{s7} and Lee, Quigg, and Thacker \cite{s8}
that $M_H$ could be bounded by requiring the interaction of
Higgs sector of SM being weak and
by using a constraint on the magnitude of partial-wave
scattering amplitude arisen from S-matrix being unitary.
They obtained the tree-level bound $M_H\sim 1.0TeV$.
Later, their analysis has been extended to the levels of one loop and
two loops by several authors \cite{s9}\cite{s10}, but the result
has been shown in a different way that the coupling starts becoming strong
around $M_H\geq 516\sim 550 GeV$, i.e. the
`unitary constraint' on $M_H$ is to be interpreted
in terms of a transition of the coupling from weak to strong,
rather than as upper bounds on the Higgs mass.

In this paper we are interested in a quite heavy Higgs boson,
$M_H>2 m_t$. In the considering case, the decay mode, Higgs boson
into a heavy $t{\bar t}$ pair, may become dominant.
Therefore a careful study of this decay mode is requested and
higher order corrections should be considered.
The QCD corrections to one loop order of the process have been computed
precisely, and to sum up the corrections up to all orders under
the leading-logarithm approximation as well, has been also made
in literature\cite{s11}.
At the first sight the EW corrections,
screened by the comparatively large QCD corrections, can not be important.
However, since the Higgs sector interaction becomes stronger as
the Higgs becomes heavier, in the considering case
the EW corrections may not be ignored, thus should be taken into
account seriously. Hence it becomes interesting to see how
the magnitudes of the one-loop EW corrections depend on $M_H$
even just in the region we are focusing here,
and whether these magnitudes suggest the break down of unitary condition.

\vskip 5mm
\noindent
{\Large{\bf 2. Calculations}}
\vskip 5mm

The lowest-order width of the decay $H\rightarrow t{\bar t}$,
corresponding to the diagram Fig.1(a), is denoted
\begin{equation}
\label{e10}
\Gamma_0=\frac {3\alpha M_Z^2 m_t^2 M_H} {8(M_Z^2-M_W^2)M_W^2} \beta^3
\end{equation}
where $\beta=(1-4 m_t^2/M_H^2)^{1/2}$ is the
usual kinematic factor.

The EW corrections to $H\rightarrow t{\bar t}$ arise from
the digrams of Fig.1-3, including counterterms.

We calculated these diagrams in the 't Hooft-Feynman gauge,
regulated all the ultraviolet and infrared divergences
by calculating in $4-2\epsilon$ dimensions,
and adopted the on-mass-shell renormalization scheme
in which the fine-structure constant $\alpha$ and the
physical masses are chosen to be the renormalized parameters.
The finite parts of the counterterms are fixed by the renormalization
conditions that the fermion propagators have poles at their masses.

With the above scheme,
and by defining the renormalization constants as below:
$$
e_0=e+\delta e,
$$
$$
M_{Z0}^2=M_Z^2+\delta M_Z^2,
$$
$$
M_{W0}^2=M_W^2+\delta M_W^2,
$$
$$
m_{t0}=m_t+\delta m_t,
$$
$$
H_0=Z_H^{\frac{1}{2}}H=(1+\delta Z_H)^{\frac{1}{2}}H,
$$
$$
\psi_{t0}=Z_t^{\frac{1}{2}}\psi_t
=(1+\delta Z_t^R \omega_+ +\delta Z_t^L \omega_-)^{\frac{1}{2}}\psi_t ,
$$
where $\omega_{\pm}={\frac{1}{2}}(1\pm \gamma_5)$,
the contribution from Fig.1 can be expressed as
\begin{equation}
\label{e1}
M_1={\bar u}(p_1)(G+\delta G)v(p_2)
=G(1+{\frac{\delta G}{G}}){\bar u}(p_1)v(p_2) ,
\end{equation}
where
\begin{equation}
G=-{\frac{ie}{2s_W}}{\frac{m_t}{M_W}} ,
\end{equation}
\begin{equation}
{\frac{\delta G}{G}}=
{\frac{\delta e}{e}}-
{\frac{1}{2}}{\frac{c_W^2}{s_W^2}}
({\frac{\delta M_Z^2}{M_Z^2}}-{\frac{\delta M_W^2}{M_W^2}})-
{\frac{1}{2}}{\frac{\delta M_W^2}{M_W^2}}+
{\frac{\delta m_t}{m_t}} .
\end{equation}
The contribution from Fig.2 can be expressed as
\begin{equation}
\label{e2}
M_2=G(\delta Z^{(v)}{\bar u}(p_1)v(p_2)
+\delta Z^{(a)}{\bar u}(p_1)\gamma_5 v(p_2)) ,
\end{equation}
where
\begin{equation}
\delta Z^{(v)}={\frac{1}{2}}(\delta Z_t^R+\delta Z_t^L)+
{\frac{1}{2}}\delta Z_H ,
\end{equation}
\begin{equation}
\delta Z^{(a)}={\frac{1}{2}}(\delta Z_t^R-\delta Z_t^L)+\cdots ,
\end{equation}
where `$\cdots$' means the contribution from Fig.2(e,f).
The contribution of Fig.2(g) turns out to be zero due to the gauge invariant.
The contribution from Fig.3 can be expressed as
\begin{equation}
\label{e3}
M_3={\bar u}(p_1)G_V v(p_2)=
G_V^{(v)}{\bar u}(p_1) v(p_2)+
G_V^{(a)}{\bar u}(p_1)\gamma_5 v(p_2) .
\end{equation}

The total renormalized amplitude may
be obtained by summing Eqs.(\ref{e1}, \ref{e2}, (\ref{e3}).
The renormalized decay rate up to next order $\alpha$,
i.e. one loop level of EW corrections, and
that of a real photon emitting as Fig.4,
is given by the following formula:
$$
\Gamma=\displaystyle \frac{\vert M_1+M_2+M_3 \vert^2}{\vert \bar{u}(p_1)
G v(p_2)\vert^2} \cdot
\Gamma_0 +
\Gamma_{rad}
$$
\begin{equation}
\label{e4}
=\displaystyle[ 1+2Re({\frac{\delta G}{G}}+
\delta Z^{(v)}+{\frac{G_V^{(v)}}{G}})] \cdot \Gamma_0+ \Gamma_{rad} .
\end{equation}
Here the term ${\frac{G_V^{(v)}}{G}}$ is due to the contribution
of the so-called `weak' vertex loop corrections shown in Fig.3, except the
virtual photonic loop diagram Fig.3 (1),
although the photonic loop contribution
to top self-energy is also subtracted in counter term
$\delta Z^{(v)}$. All the
photonic loop contribution is treated together with
that of a real photon emitting (shown in Fig.4), and as a common
`QED correction' we denote it as $\Gamma_{rad}$. Calculations show that
\arraycolsep1.5pt
\begin{eqnarray*}
{\frac{\delta G}{G}}&=&
{\frac{\alpha}{4\pi}}({\frac{1}{4\pi\mu^2}})^{\epsilon} \{
[{\frac{17}{24c_W^2}}+{\frac{9}{8s_W^2}}-\\
&&{\frac{1}{4c_W^2 s_W^2 M_Z^2}}
\sum\limits_{j=1}^{3}(m_{l,j}^2+3 m_{u,j}^2+3 m_{d,j}^2)-
{\frac{1}{8c_W^2 s_W^2 M_Z^2}}
\sum\limits_{j=1}^{3}(K_{3j}K_{j3}^{\dag} m_{d,j}^2)
\end{eqnarray*}
\begin{equation}
+{\frac{1}{2c_W^2 s_W^2 M_Z^2}}
\sum\limits_{j=1}^{3}(K_{3j}K_{j3}^{\dag} m_{d,j} m_t)]
\Delta(\epsilon)_{UV}+finite\ part\} ,
\end{equation}

\begin{eqnarray*}
\delta Z^{(v)}&=&
{\frac{\alpha}{4\pi}}({\frac{1}{4\pi\mu^2}})^{\epsilon} \{
[-{\frac{1}{72c_W^2}}-{\frac{3}{8s_W^2}}+
{\frac{1}{4c_W^2 s_W^2 M_Z^2}}
\sum\limits_{j=1}^{3}(m_{l,j}^2+3 o_{u,j}^2+3 m_{d,j}^2)
\end{eqnarray*}
\begin{equation}
+{\frac{1}{8c_W^2 s_W^2 M_Z^2}}
\sum\limits_{j=1}^{3}(K_{3j}K_{j3}^{\dag} m_{d,j}^2)
]\Delta(\epsilon)_{UV}
-{\frac{8}{9}}\Delta(\epsilon)_{IR}
+finite\ part\} ,
\end{equation}

\begin{eqnarray*}
{\frac{G_V^{(v)}}{G}}&=&
{\frac{\alpha}{4\pi}}({\frac{1}{4\pi\mu^2}})^{\epsilon}
\{[-{\frac{25}{36c_W^2}}-{\frac{3}{4s_W^2}}-
{\frac{1}{2c_W^2 s_W^2 M_Z^2}}
\sum\limits_{j=1}^{3}(K_{3j}K_{j3}^{\dag} m_{d,j} m_t)
]\Delta(\epsilon)_{UV}
\end{eqnarray*}
\begin{equation}
+{\frac{4L}{9}}{\frac{1+\beta^2}{\beta}}\Delta(\epsilon)_{IR}
+finite\ part\} ,
\end{equation}
where $L=\ln ((1+\beta)/(1-\beta))$ and $j$ is
the generation number.
We have distinguished the divergences of ultraviolet
origin from those of infrared origin with a subscript,
and the long expressions of the finite parts have been suppressed.

The most important contribution comes from $\delta Z_H$,
which is calculated from Fig.2(c,d):

\begin{eqnarray*}
\delta Z_H &=&
\frac{\alpha}{4\pi}(\frac{1}{4\pi\mu^2})^{\epsilon}
\{
\sum\limits_{j=1}^{3} \frac{m_{l,j}^2}{2s_W^2 M_W^2}
[\Delta(\epsilon)+F(M_H^2,m_{l,j},m_{l,j})+M_H^2 F'(M_H^2,m_{l,j},m_{l,j})] \\
&+&\sum\limits_{j=1}^{3} \frac{3m_{u,j}^2}{2s_W^2 M_W^2}
[\Delta(\epsilon)+F(M_H^2,m_{u,j},m_{u,j})+M_H^2 F'(M_H^2,m_{u,j},m_{u,j})] \\
&+&\sum\limits_{j=1}^{3} \frac{3m_{d,j}^2}{2s_W^2 M_W^2}
[\Delta(\epsilon)+F(M_H^2,m_{d,j},m_{d,j})+M_H^2 F'(M_H^2,m_{d,j},m_{d,j})] \\
&-& \frac{1}{s_W^2} [ \Delta(\epsilon)+F(M_H^2,M_W,M_W)-
(3M_W^2-M_H^2+ \frac{M_H^4}{4M_W^2} )F'(M_H^2,M_W,M_W) ] \\
&-& \frac{1}{2s_W^2 c_W^2} [ \Delta(\epsilon)+F(M_H^2,M_Z,M_Z)-
(3M_Z^2-M_H^2+ \frac{M_H^4}{4M_Z^2} )F'(M_H^2,M_Z,M_Z) ]
\end{eqnarray*}
\begin{equation}
+{\frac{9}{8s_W^2}}{\frac{M_H^4}{M_W^2}}F'(M_H^2,M_H,M_H)
+O(\epsilon) \} ,
\end{equation}
where
\begin{equation}
F(s,m_0,m_1)=\int_0^1 dx \ln {\frac{x^2 s-x(s-m_0^2+m_1^2)+m_1^2-i\epsilon}
{\mu^2}}
\end{equation}
and
\begin{equation}
F'(s,m_0,m_1)= {\frac{\partial F(s,m_0,m_1)}{\partial s}} .
\end{equation}

To get rid of the infrared divergences and
complete the calculation of the total decay rate,
we calculated the decay and the radiative decay $H \to t\bar t \gamma$
together as mentioned above.
To match the common calculations of QED\cite{s11}, in which a infrared
regularization by introducing a tiny mass of photon
is adopted and being different here, we divide artificially
the photon into hard and soft with a criterion energy $\lambda$,
so that it is easy to complete the match of ours and theirs
at the energy of photon.
The $\lambda$ dependence cancels finally if summing the hard and soft
contributions as well as those of the two concerned processes, which
all are matched well at the photon energy $\lambda$.
The result summed as above is \cite{s11}:
\begin{eqnarray*}
\Gamma_{rad}/\Gamma_0 &=&
{\frac{\alpha}{4\pi}}({\frac{1}{4\pi\mu^2}})^{\epsilon} \{
({\frac{16}{9}}-{\frac{8L}{9}}{\frac{1+\beta^2}{\beta}})\Delta(\epsilon)_{IR}
+{\frac{64}{9}}-{\frac{2}{3}}{\frac{1+\beta^2}{\beta^2}}+\\
&&{\frac{64}{9}}\ln {\frac{m_t}{M_H}}-{\frac{64}{9}}\ln \beta+
({\frac{8}{3\beta}}+{\frac{(1-\beta^2)^2}{3\beta^3}})L+
{\frac{8}{9}} {\frac{1+\beta^2}{\beta}}\cdot \\
&&[-{\frac{L^2}{2}}+4L\ln{\frac{1+\beta}{2}}-2L\ln\beta-
2L\ln{\frac{1+\beta}{2}}\ln{\frac{1-\beta}{2}}+
\end{eqnarray*}
\begin{equation}
\label{e100}
6 Sp({\frac{1+\beta}{1-\beta}})-4 Sp({\frac{1-\beta}{2}})-
{\frac{2\pi^2}{3}}]\} ,
\end{equation}
where $Sp(x)=-\int_0^x (dt/t)ln(1-t)$ is the Spence function.
{}From Eqs.(\ref{e4}-\ref{e100}), one may see that the
ultraviolet and infrared divergences have canceled as they should.

\vskip 5mm
\noindent
{\Large{\bf 3. Results and Discussions}}
\vskip 5mm

For the numerical evaluation, we use following the set of independent
parameters which are currently known with the highest
experimental accuracy\cite{pdg}:
$$
\begin{array}{lll}
\alpha=1/137.0359895, &\qquad G_F=1.16637\times 10^{-5} GeV^{-2},
&\qquad M_Z=91.175 GeV,\\
m_e=0.51099906 MeV, &\qquad m_{\mu}=0.10565839 GeV,
&\qquad m_{\tau}=1.7841 GeV, \\
m_u=41 MeV, &\qquad m_d=41 MeV, &\qquad m_s=150 MeV, \\
m_c=1.5 GeV, &\qquad m_b=4.5 GeV. &
\end{array}
$$

The masses of the light quarks are chosen such as that the
experimentally determined hadronic vacuum polarization is reproduced\cite{fj}.
The mass of W-boson is induced from these parameters, and the muon decay
measured width to the theoretical calculations
with the radiative and up to one-loop weak corrections\cite{as}.
This yields a more precise value of $M_W$, $s_W$, and further, the
lowest-order cross sections (or widths) will depend on $m_t$ and $M_H$
through $M_W$ and/or $s_W$ if the laters appear in the formulas
of the concerned processes.
In our calculation, $M_W$ is determined:
\begin{equation}
\label{e12}
M_W^2(1-{\frac{M_W^2}{M_Z^2}})={\frac{\pi\alpha}{\sqrt{2}G_F (1-\Delta r)}} ,
\end{equation}
where $\Delta r$ involves the weak and radiative corrections to the
muon decay\cite{as}. According to Ref.\cite{wj},
\begin{equation}
\Delta r\simeq 0.0696-{\frac{3\alpha}{16\pi}}{\frac{c_W^2}{s_W^2}}
{\frac{m_t^2}{M_W^2}}+{\frac{11\alpha}{48\pi s_W^2}}
\ln{\frac{M_H^2}{M_W^2}}
\end{equation}
for $m_t^2\gg M_W^2$ and $M_H^2\gg M_W^2$.

In Fig.5, we present the EW corrections to the decay rate
$\Gamma /\Gamma_0 (H\rightarrow t{\bar t})$ for
a 130-GeV top-quark,
and compare it with the separated QED corrections
and the QCD corrections taken from Ref\cite{s11}.\footnote{ Throughout
the paper, various top masses are still discussed not only for keeping
the original aspect of the paper as much as possible in present version
but also for comparing our calculations with others' calculations
directly, although
a fresh top mass value has been presented by CDF collaboration.}

The EW corrections are in opposite direction to the QCD corrections.
The EW corrections have a tendency to broaden the decay
width being sizable,
due to large $m_{t}$ and $M_{H}$ in our discussion.
In the range of Higgs mass $M_H\leq 1. TeV$, the EW and QCD corrections
are comparable in magnitude and nearly cancel each other. Finally
the total corrections just reduce to a few percent.
However for the range $M_H>1. TeV$, the EW corrections exceed the QCD
corrections
in magnitude gradually, because some diagrams
of the EW corrections involve the self coupling of the Higgs, and it is
proportional to $M_H^2$ ( the vacuum expectation value
of the EW spontaneous breaking is fixed in SM at all ),
so the EW corrections arise rapidly, while in contrary, of the QCD corrections,
there is no such a factor at all.
In fact, the large magnitude of EW corrections at $M_H>1. TeV$ only
suggests that the `weak' interaction,
as pointed out just now, has indeed become strong,
and means the perturbation calculation is invalid. We calculated
the corrections out and put the numerical results in Fig.5, here
even those range are involved out off the calculations
being valid, just only for a reference.
Intuitively one may understand the results in another way
that the fermions and the Higgs will not decouple even when the
masses of them are approaching to infinity,
due to the fact that the Higgs itself coupling
is proportional to the Higgs mass squared and the one of Higgs to
fermion pair proportional to the fermion mass respectively,
while the QCD corrections are always controlled by the decouple theorem,
and behave in the manner dominated by
logarithmic terms such as $\ln{\frac{m_t}{M_W}}$.

Note that there would be a spike at the threshold $M_H=2 m_t$
if one just stopped at one loop level corrections.
It arises from the occurrence of $\beta^{-1}$
in the QED one loop correction.
At a place very close to the threshold, one would clearly have to add all
terms in the form as $(\alpha/\beta)^n$, so as to obtain
a reliable result. Schwinger has shown that this
summation is equivalent to the use of an $1/r$ potential
to describe the electric interactions between the quark and
the antiquark\cite{js}. Actually the Eq.(16) can not be used in
the Higgs mass range near the $t\bar{t}$ threshold. In
the case near the threshold, to pursue a correct result one must
do whole perturbative QED calculation and sum up such leading terms.
Being not what we are interested in here, we avoid the problem
and present the results not so close to the threshold.
The same thing happened in the QCD one-loop
calculation. The QCD curve, we present in Fig.5, is the
leading-logarithm approximation results which have been
obtained with $\beta\rightarrow 1$ too.

In order to compare the calculations here with the earlier ones
we plotted the various decay rates of
$H\rightarrow t{\bar t}$ for a 100-GeV top-quark,
a 130-GeV top-quark and a 150-GeV top-quark respectively
in Fig.6. In the figure
the decay rates obtained from a tree level,
those from one-loop EW corrections
concerned only and those from both the EW (up to one-loop)
and the QCD (up to LLA) corrections concerned
are presented with the dotted, the dashed and the solid curves respectively.
Being a bone of the paper,
the corresponding
`weak' one-loop corrections $\delta \Gamma$ to
the tree level width of the process $H \rightarrow t\bar{t}$
as a function of Higgs mass are presented in Fig.7
for various possible top masses, especially, the fresh one.\cite{s21}

For comparison, we would like to discuss another parametrization
scheme of the renormalization,
called the $G_F$-scheme here, in which the lowest-order
expression is parametrized in terms of $G_F$ instead of $\alpha$.
As suggested in Ref.\cite{ge}, for the processes dominated by
mass scales larger than $M_W$, it becomes more appropriate.
In the $G_F$-scheme, $\Gamma_0^{G_{F}}$ is given by
\begin{equation}
\label{e11}
\Gamma_0^{G_F}={\frac{3}{4\pi}}G_F m_t^2 M_H\beta^3 .
\end{equation}
{}From Eqs.(\ref{e10}, Eq.\ref{e12}), and Eq.(\ref{e11}),
we have the relation between the two schemes
\begin{equation}
\Gamma_0^{G_F}=\Gamma_0/(1-\Delta r) ,
\end{equation}
\begin{equation}
\Gamma=\Gamma_0(1+\delta_{EW})=\Gamma_0^{G_F}(1+\delta_{EW}^{G_F}) .
\end{equation}
Then approximately we have
\begin{equation}
\delta_{EW}^{G_F}=\delta_{EW}-\Delta r .
\end{equation}
The EW corrections in a $G_F$-scheme $\delta_{EW}^{G_F}$ is compared
with $\delta_{EW}$ in Fig.8,
which indicates that a considerable reduction of the
EW corrections (about 7\%-8\%, when $M_{H}<1 TeV$) are obtained in the $G_F$
parametrization scheme.
\par
There are similar works in literature such as the calculations
done by A. Dabelstein, W. Hollik and B. A. Kniehl Refs.\cite{s18} \cite{s19}.
They figured out the weak corrections at
$m_{t}=100$ GeV, $150$ GeV, $200$ GeV and $250$ GeV respectively.
A careful comparison of our results with corresponding ones in references
Ref\cite{s18} and Ref.\cite{s19} was made. By
changing to the $G_{F}$-scheme,
we can find that our numerical results are complete
coincidence with
theirs, but one should note that for an accurate numerical comparison it is
necessary to use Eq.$(23)$ instead of approximate Eq.$(22)$,
especially when the value of $\delta_{W}$ becomes large.
\begin{equation}
\delta_{W}^{G_F}=\delta_{W}-\Delta r-\delta_{W}\Delta r.
\end{equation}
As the total corrections yield a reduction of a few percent in the width
when $M_{H}< 1. TeV$,
it may be feasible to test all the calculations
in future experiments, only when the
precision of the experimental measurements on the process
of $H \rightarrow t\bar{t}$ and its radiative mode
is less than few percent and if the Higgs mass
is much larger than the threshold of $H \rightarrow t\bar{t}$ decay as well.

\vskip 25mm
\noindent
{\Large{\bf Acknowledgements}}
\vskip 5mm
This work was supported in part by the National
Natural Science Foundation of China under Grants No.19275040
and the most important projects. It was also supported in
part by the Grant LWTZ-1298 of Chinese Academy of Science.

\newpage
\noindent
{\Large{\bf Appendix A. Definitions}}
\vskip 2mm
\par
We adopt the definitions of one-loop integral functions as
$$
A_{0}(m^2)=-{\frac{(2\pi\mu)^{4-D}}{i\pi^2}}\int d^D q
{\frac{1}{[q^2-m^2]}},
$$
$$
\{B_0;B_{\mu};B_{\mu\nu}\}(p^2,m_1,m_2)=
{\frac{(2\pi\mu)^{4-D}}{i\pi^2}}\int d^D q
{\frac{\{1;q_{\mu};q_{\mu\nu}\}}{[q^2-m_1^2][(q+p)^2-m_2^2]}},
$$
$$
\{C_0;C_{\mu};C_{\mu\nu};C_{\mu\nu\rho}\}(p^2,k^2,(p+k)^2,m_1,m_2,m_3)=
$$
$$
-{\frac{(2\pi\mu)^{4-D}}{i\pi^2}}\int d^D q
{\frac{\{1;q_{\mu};q_{\mu\nu};q_{\mu\nu\rho}\}}
{[q^2-m_1^2][(q+p)^2-m_2^2][(q+p+k)^2-m_3^2]}},
$$
in conjunction with the decompositions
$$
B_{\mu}=p_{\mu}B_{1},
$$
$$
B_{\mu\nu}=p_{\mu}p_{\nu}B_{21}-g_{\mu\nu}B_{22},
$$
$$
C_{\mu}=p_{\mu}C_{11}+k_{\mu}C_{12},
$$
$$
C_{\mu\nu}=p_{\mu}p_{\nu}C_{21}+k_{\mu}k_{\nu}C_{22}
+(p_{\mu}k_{\nu}+k_{\mu}p_{\nu})C_{23}-g_{\mu\nu}C_{24}.
$$
\par
The explicit analytic forms of above functions one can find
in references \cite{s20}.

\vskip 5mm
\noindent
{\Large{\bf Appendix B. Vertex corrections}}
\vskip 2mm

\par
The vertex corrections at one-loop level as shown in Fig.3 can be written
in the form
\begin{eqnarray*}
\frac{G_{V}}{G}=\sum\limits_{i=1}^{15}\Delta T_{i}.
\end{eqnarray*}
where $\Delta T_{i}$ represent the vertex diagrams (i) of Fig.3. The
analytic formulae are listed as the follows:
\begin{eqnarray*}
\Delta T_{1}&=&\frac{\alpha}{9\pi}\{DB_{0}(M_{H}^{2},m_{t},m_{t})+
[2M_{H}^{2}(C_{0}+C_{11})-D\lambda^2C_{0}\\[2mm]
&-&8m_{t}^{2}(C_{0}+\frac{3}{2}C_{11} -C_{12})+
2Dm_{t}^{2}(C_{11}-2C_{12})]
(m_{t}^2,M_{H}^2,m_{t}^2,\lambda,m_{t},m_{t})\}, \\[4mm]
\Delta T_{2}&=&\frac{\alpha}{144\pi s^{2}_{W}c^{2}_{W}}\{4Ds_{W}^{2}(-3+
4s^{2}_{W})B_{0}(M_{H}^{2},m_{t},m_{t})\\ [2mm]
&+&[8s_{W}^{2}(4s_{W}^{2}-3)[M_{H}^2
(C_{0}+C_{11})-\frac{D}{2}M_{Z}^{2}C_{0}+
m_{t}^{2}(DC_{11}-4C_{0}-6C_{11})]\\[2mm]
&+&(2-D)m_{t}^2[9(2C_{12}-C_{11})+16s_{W}^2(4s_{W}^2-3)C_{12}]]
(m_{t}^2,M_{H}^2,m_{t}^2,M_{Z},m_{t},m_{t})\}, \\[4mm]
\Delta T_{3}&=&\frac{\alpha |K_{33}|^{2}m_{b}^{2}(D-2)}
{8\pi s^{2}_{W}}[C_{11}-2C_{12}]
(m_{t}^2,M_{H}^2,m_{t}^2,M_{W},m_{b},m_{b}),   \\[4mm]
\Delta T_{4}&=&\frac{\alpha m_{t}^{2}}{16\pi s^{2}_{W}M_{W}^{2}}
\{-B_{0}(M_{H}^{2},m_{t},m_{t})\\[2mm]
&+&[M_{H}^{2}C_{0}+2m_{t}^{2}(C_{11}-2C_{12})]
(m_{t}^2,M_{H}^2,m_{t}^2,M_{H},m_{t},m_{t})\},  \\[4mm]
\Delta T_{5}&=&\frac{\alpha m_{t}^{2}}{16\pi s^{2}_{W}M_{W}^{2}}
\{B_{0}(M_{H}^{2},m_{t},m_{t})\\[2mm]
&-&[M_{Z}^{2}C_{0}-2m_{t}^{2}(C_{11}-2C_{12})]
(m_{t}^2,M_{H}^2,m_{t}^2,M_{Z},m_{t},m_{t})\},    \\[4mm]
\Delta T_{6}&=&-\frac{\alpha m_{b}^{2}|K_{33}|^{2}}{8\pi s^{2}_{W}M_{W}^{2}}
\{-B_{0}(M_{H}^{2},m_{b},m_{b})+[m_{b}^{2}(C_{0}-C_{11}+2C_{12})\\[2mm]
&-& m_{t}^{2}(C_{0}+C_{11}-2C_{12})+M_{W}^{2}C_{0}]
(m_{t}^2,M_{H}^2,m_{t}^2,M_{W},m_{b},m_{b})\},        \\[4mm]
\Delta T_{7}&=&\frac{3\alpha M^{2}_{H}m^{2}_{t}}{16\pi
s^{2}_{W}M^{2}_{W}} [C_{0}-C_{11}+2C_{12}]
(m_{t}^2,M_{H}^2,m_{t}^2,m_{t},M_{H},M_{H}), \\[4mm]
\Delta T_{8}&=&-\frac{\alpha M^{2}_{H} m^{2}_{t}}{16\pi s^{2}_{W}M^{2}_{W}}
[C_{0}+C_{11}-2C_{12}]
(m_{t}^2,M_{H}^2,m_{t}^2,m_{t},M_{Z},M_{Z}), \\[4mm]
\end{eqnarray*}
\begin{eqnarray*}
\Delta T_{9}&=&-\frac{\alpha M^{2}_{H}}{16\pi s^{2}_{W}M^{2}_{W}}
[2 m_{b}^{2}(C_{0}+\frac{C_{11}}{2}-C_{12})\\[2mm]
&+&m_{t}^{2}(C_{11}-2C_{12})]
(m_{t}^2,M_{H}^2,m_{t}^2,m_{b},M_{W},M_{W}), \\
\Delta T_{10}&=&
-\frac{\alpha M^{2}_{Z}}{144 \pi c^{2}_{W} s^{2}_{W}}[ 8D s^{2}_{W}(-3
+4s^{2}_{W})C_{0}+(2-D)(9-24s^{2}_{W}+32s^{4}_{W})\\[2mm]
&&(2C_{12}-C_{11})](m_{t}^2,M_{H}^2,m_{t}^2,m_{t},M_{Z},M_{Z}), \\[4mm]
\Delta T_{11}&=&
-\frac{\alpha M^{2}_{W}(D-2)}{8 \pi s^{2}_{W}} (C_{11}-2C_{12})
(m_{t}^2,M_{H}^2,m_{t}^2,m_{b},M_{W},M_{W}), \\[4mm]
\Delta T_{12}&=&-\frac{\alpha}{32\pi c^{2}_{W} s^{2}_{W}}\{-B_{0}(M^{2}_{H},
 M_{Z},M_{Z})+[4m^{2}_{t}(C_{0}+C_{11})\\[2mm]
&-& 2M^{2}_{H}C_{12})]
(m_{t}^2,M_{H}^2,m_{t}^2,m_{t},M_{Z},M_{Z})\}, \\[4mm]
\Delta T_{13}&=&-\frac{\alpha}{16\pi s^{2}_{W}}\{-B_{0}(M^{2}_{H},
M_{W},M_{W})+[2m^{2}_{b}(2C_{0}+{\frac {C_{11}}{2}}-C_{12})\\[2mm]
&+& m^{2}_{t}(3C_{11}-2C_{12})-2M_{H}^2C_{12}]
(m_{t}^2,M_{H}^2,m_{t}^2,m_{b},M_{W},M_{W})\}, \\[4mm]
\Delta T_{14}&=&
\frac{\alpha}{32\pi c^{2}_{W} s^{2}_{W}} \{B_{0}(M^{2}_{H},M_{Z},M_{Z})+
[2m^{2}_{t}(C_{0}-3C_{11}+2C_{12})\\[2mm]
&+& 2M^{2}_{H}(C_{11}-C_{12})]
(m_{t}^2,M_{H}^2,m_{t}^2,m_{t},M_{Z},M_{Z})\}, \\[4mm]
\Delta T_{15}&=&-\frac{\alpha}{16\pi s^{2}_{W}}\{-B_{0}(M^{2}_{H},
M_{W},M_{W})-[2m^{2}_{b}(C_{0}+C_{12}-{\frac{C_{11}}{2}})\\[2mm]
&-&2M^{2}_{H}(C_{12}-C_{11})-2m^{2}_{t}({\frac{5C_{11}}{2}}-C_{12})]
(m_{t}^2,M_{H}^2,m_{t}^2,m_{b},M_{W},M_{W})\}.
\end{eqnarray*}
In above formulae $\lambda$ means a small photon mass
and $D=4-2 \epsilon$. The arguments of C-functions are written at the end
of formulae in parenthesis.

\newpage
\noindent
\centerline{\Large \bf Figure Captions}
\vskip 5mm
\noindent
{\bf Fig.1} \hspace{2mm} (a) The tree level diagrams,
and (b) the vertex counterterms diagrams.
\vskip 2mm
\noindent
{\bf Fig.2} \hspace{2mm} The diagrams with one loop on the external lines, each
shadowed circle includes all the possible one-loops and
the corresponding counterterms.
\vskip 2mm
\noindent
{\bf Fig.3} \hspace{2mm} The one-loop vertex diagrams.
\vskip 2mm
\noindent
{\bf Fig.4} \hspace{2mm} The diagrams with one real photon emission.
\vskip 2mm
\noindent
{\bf Fig.5} \hspace{2mm} The EW corrections to the decay rate of
$H\rightarrow t{\bar t}$ for a 130-GeV top-quark,
the pure QED corrections,
and the QCD corrections\cite{s11}.
\vskip 2mm
\noindent
{\bf Fig.6} \hspace{2mm} The decay rate of
$H\rightarrow t{\bar t}$ for a 100-GeV top-quark,
a 130-GeV top-quark and a 150-GeV top-quark.
The dotted lines correspond to the decay rates with
the various top masses at the tree level order,
the long-dashed lines to those at the one loop order of
EW corrections, and the solid lines to those at the order of
EW (one loop) and QCD (LLA) corrections. The QCD corrections
are quoted from Ref\cite{s11}.
\vskip 2mm
\noindent
{\bf Fig.7} \hspace{2mm} The weak corrections
to the decay width of $H \rightarrow t\bar{t}$
as a function of Higgs mass, with various top masses
$m_{t}=100 GeV, 130 GeV, 150 GeV$ and $174 GeV$ respectively
and in $\alpha$-scheme.
\vskip 2mm
\noindent
{\bf Fig.8} \hspace{2mm} The EW corrections in a $G_F$
parametrization scheme $\delta_{EW}^{G_F}$ compared
with $\delta_{EW}$.

\end{large}

\newpage
\begin{thebibliography}{s20}
\bibitem{s1} J. F. Gunion {\sl et. al.}, The Higgs Hunter's Guide
             (Addison-Wesley, Reading, MA, 1990).
\bibitem{s2} J. L. Rosner, Rev. of Mod. Phys. {\bf 64}, 1151(1992).
\bibitem{s3} M. Davier, in {\sl Proceedings of the Joint International
             Lepton-Photon Symposium and Europhysics Conference on
             High Energy Physics}. Geneva, Switzerland, 1991, edited
             by S. Hegarty, K. Potter, and E. Quereigh (World Scientific,
             Singapore, 1992).
\bibitem{s4} W. Hollik, Talk at the ``XVI international symposium
             on Lepton-Photon Interactions'', Cornell University,
             Ithaca, New York, 10-15 August, 1993.
\bibitem{s5} For recent analysis see
             P. Langacker and M. Luo, Phys. Rev. D {\bf 44}, 817(1991),
             and references therein.
\bibitem{s6} M. Veltman, Acta Phys. Pol. B {\bf 8}, 475(1977).
\bibitem{s7} Dicus and Mathur, Phys. Rev. D {\bf 7}, 3111(1973).
\bibitem{s8} B. W. Lee, C. Quigg, and H. B. Thacker,
             Phys. Rev. Letts {\bf 38}, 883(1977);
             B. W. Lee, C. Quigg, and H. B. Thacker,
             Phys. Rev. D {\bf 16}, 1519(1977).
\bibitem{s9} For example, see
             L. Durand, J. M. Johnson, and J. L. Lopez,
             Phys. Rev. Letts {\bf 64}, 1215(1990);
             L. Durand, J. M. Johnson, and J. L. Lopez,
             Phys. Rev. D {\bf 45}, 3112(1992).
\bibitem{s10}P. N. Maher, L. Durand, and K. Riesselmann,
             Phys. Rev. D {\bf 48}, 1061(1993);
             L. Durand, P. N. Maher, and K. Riesselmann,
             Phys. Rev. D {\bf 48}, 1084(1993).
\bibitem{s11}E. Braaten and J. P. Leveille,
             Phys. Rev. D {\bf 22}, 715(1980);
             A pertinent review can be found in
             P. J. Franzini and P. Taxil,
             `Z Physics at LEP1', vol.2, P.59 (CERN yellow
             report 89-08, edited by G. Altarelli, R. Kleiss
             and C. Verzegnassi, 1989).
\bibitem{pdg}Particle Data Group, Phys. Letts. B {\bf 239}, 1(1990).
\bibitem{fj} F. Jegerlehner, in Proc. 1990 Theoretical Advanced Study
             Institute in Elementary Particle Physics, edited by
             M. Cveti\v{c} and P. Langacker (World Scientific,
             Singapore, 1991), p.476.
\bibitem{as} A. Sirlin, Phys. Rev. D{\bf 22}, 971(1980).
\bibitem{wj} W. J. Marciano and Z. Parsa, Annu. Rev. Nucl. Sci.
             {\bf 36}, 171(1986).
\bibitem{js} J. Schwinger, {\sl Particles, Sources and Fields},
             vol.II (Addison-Wesley, Reading, MA, 1973).
\bibitem{ge} G. Eilam, P. R. Mendel, R. Migeron and A. Soni,
             Phys. Rev. Letts. {\bf 66}, 3105(1991).
\bibitem{s18}A. Dabelstein and W. Hollik, Z. Phys. C {\bf 53}, 507(1992).
\bibitem{s19}B. A. Kniehl, Nucl. Phys. B{\bf 376}, 3(1992).
\bibitem{s20}M. Bohm, W. Hollik, H. Spiesberger, Fortschr. Phys. {\bf 34},
             687(1986);
             A. Denner, Fortschr.Phys. {\bf 41}, 307(1993);
             B. A. Kniehl, DESY 93-069, August 1993.
\bibitem{s21}CDF Collaboration, {\bf FERMILAB-PUB-94/097-E}.
\end{thebibliography}
\end{document}